\documentclass{article}

\usepackage{arxivsimplified}

\usepackage[utf8]{inputenc} 
\usepackage[T1]{fontenc}    
\usepackage{hyperref}       
\usepackage{url}            
\usepackage{booktabs}       
\usepackage{amsfonts}       
\usepackage{nicefrac}       
\usepackage{microtype}      
\usepackage{lipsum}

\usepackage{graphicx}
\graphicspath{{figs_pdf/}}
\usepackage{subfigure}
\usepackage{cite}
\usepackage[dvipsnames]{xcolor}
\usepackage{amsmath}
\usepackage{blindtext}
\usepackage{amsfonts}
\usepackage{multirow}
\usepackage{array}
\usepackage{mathtools}
\usepackage{listings}
\usepackage{xcolor}
\definecolor{codegreen}{rgb}{0,0.6,0}
\definecolor{codegray}{rgb}{0.5,0.5,0.5}
\definecolor{codepurple}{rgb}{0.58,0,0.82}
\definecolor{backcolour}{rgb}{0.95,0.95,0.92}
\usepackage{color}
\usepackage{xcolor}
\definecolor{rev1}{rgb}{0,0,0}
\usepackage{esvect}
\usepackage{tabularx}
\newcolumntype{L}[1]{>{\raggedright\arraybackslash}p{#1}}
\newcolumntype{C}[1]{>{\centering\arraybackslash}p{#1}}
\newcolumntype{R}[1]{>{\raggedleft\arraybackslash}p{#1}}

\DeclareMathOperator{\argmin}{arg\,min} 

\usepackage{algorithm}
\usepackage{algorithmic}
\usepackage{enumitem}
\setlist[itemize]{leftmargin=*}

\lstdefinestyle{mystyle}{
    backgroundcolor=\color{backcolour},   
    commentstyle=\color{codegreen},
    keywordstyle=\color{magenta},
    numberstyle=\tiny\color{codegray},
    stringstyle=\color{codepurple},
    basicstyle=\ttfamily\footnotesize,
    breakatwhitespace=false,         
    breaklines=true,                 
    captionpos=b,                    
    keepspaces=true,                 
    numbers=left,                    
    numbersep=5pt,                  
    showspaces=false,                
    showstringspaces=false,
    showtabs=false,                  
    tabsize=2
}
 
\usepackage{scalerel,stackengine}
\stackMath
\newcommand\reallywidehat[1]{%
\savestack{\tmpbox}{\stretchto{%
  \scaleto{%
    \scalerel*[\widthof{\ensuremath{#1}}]{\kern-.6pt\bigwedge\kern-.6pt}%
    {\rule[-\textheight/2]{1ex}{\textheight}}
  }{\textheight}%
}{0.5ex}}%
\stackon[1pt]{#1}{\tmpbox}%
}

\title{Model fusion with physics-guided machine learning}

\author{
    Suraj Pawar  \\
    School of Mechanical \& Aerospace Engineering,\\
    Oklahoma State University, \\
    Stillwater, Oklahoma - 74078, USA.\\
    \texttt{supawar@okstate.edu} \\
    \And
    Omer San \\
    School of Mechanical \& Aerospace Engineering,\\
    Oklahoma State University, \\
    Stillwater, Oklahoma - 74078, USA.\\
    \texttt{osan@okstate.edu} \\
    \And
    Aditya Nair \\
    Department of Mechanical Engineering, \\
    University of Nevada, \\
    Reno, NV 89557, USA
    \And
    Adil Rasheed \\
	Department of Engineering Cybernetics, \\
	Norwegian University of Science and Technology, \\
	7465 Trondheim, Norway. \\
    Department of Mathematics and Cybernetics, \\
    SINTEF Digital, \\
    7034 Trondheim, Norway. \\
    \AND
    Trond Kvamsdal \\
    Department of Mathematics and Cybernetics, \\
    SINTEF Digital, \\
    7034 Trondheim, Norway. \\
    Department of Mathematical Sciences, \\
    Norwegian University of Science and Technology, \\
    7491 Trondheim, Norway.
}

\begin{document}
\maketitle

\begin{abstract}
The unprecedented amount of data generated from experiments, field observations, and large-scale numerical simulations at a wide range of spatio-temporal scales have enabled the rapid advancement of data-driven and especially deep learning models in the field of fluid mechanics. Although these methods are proven successful for many applications, there is a grand challenge of improving their \emph{generalizability}. This is particularly essential when data-driven models are employed within outer-loop applications like optimization. In this work, we put forth a physics-guided machine learning (PGML) framework that leverages the interpretable physics-based model with a deep learning model.  The PGML framework is capable of enhancing the generalizability of data-driven models and effectively protect against or inform about the inaccurate predictions resulting from extrapolation. We apply the PGML framework as a novel model fusion approach combining the physics-based Galerkin projection model and long-short term memory (LSTM) network for parametric model order reduction of fluid flows. We demonstrate the improved generalizability of the PGML framework against a purely data-driven approach through the injection of physics features into intermediate LSTM layers. Our quantitative analysis shows that the overall model uncertainty can be reduced through the PGML approach especially for test data coming from a distribution different than the training data. Moreover, we demonstrate that our approach can be used as an inverse diagnostic tool providing a confidence score associated with models and observations. The proposed framework also allows for multi-fidelity computing by making use of low-fidelity models in the online deployment of quantified data-driven models.     
\end{abstract}


\section{Introduction}

Data-driven approaches are emerging as the new paradigm in computational modeling of various problems in different branches of science and engineering, including fluid dynamics \cite{brunton2020machine,brenner2019perspective}. The universal approximation capability of neural networks  \cite{hornik1989multilayerua} makes them the powerful algorithm for complicated problems like turbulence closure modeling \cite{duraisamy2019turbulence}, spatio-temporal super-resolution \cite{fukami2021machine}, state-estimation \cite{nair2020leveraging}, and nonlinear model order reduction \cite{lee2020model}. One of the key issues with deep learning is that they exhibit poor \emph{generalizability}, i.e., they produce inaccurate prediction when the test data is from a distribution far from the training data. This adversely affects the trustworthiness of neural networks for scientific applications and embedding the known physics to enhance the generalizability is a challenge and opportunity for data-driven algorithm developments \cite{karpatne2017theory,reichstein2019deep,rasheed2020digital,kashinath2021physics}.

While many techniques have sought to enforce physics into data-driven models, such as regularizing the neural network with the governing equations or statistical constraints \cite{raissi2019physics,wu2020enforcing,geneva2020modeling}, imposing conservation of physical quantities directly into neural network \cite{beucler2019achieving,mohan2020embedding}, embedding invariance property into neural network architecture \cite{ling2016reynolds,zanna2020data}, and incorporating inductive biases to respect exact conservation laws \cite{greydanus2019hamiltonian}, they offer many possibilities to fuse domain knowledge to improve the generalizability and explainability of these models \cite{kashinath2021physics}. In this work, we propose the physics-guided machine learning (PGML) framework that can answer the question of how to hybridize data-driven (i.e., non-intrusive) and first-principles (i.e., intrusive) approaches that are often utilized in fluid dynamics applications. The PGML framework is based on a composite modeling strategy where the physics-based model is integrated within the neural network architecture leading to a more generalizable learning engine.  

\textcolor{rev1}{The high-dimensional and multiscale nature of fluid flows makes the numerical discretization methods computationally infeasible for many practical applications. There are several techniques that aim at at constructing the lower-order representation of high-dimensional systems that can capture the essential features and are computationally order of magnitudes faster than full order model \cite{rowley2017model,benner2015survey,taira2019modal}}.
These reduced order models (ROMs) are particularly appealing for outer-loop applications, such as data assimilation \cite{cao2007reduced,cstefuanescu2015pod,meldi2017reduced}, uncertainty quantification \cite{guzzetti2020propagating}, model predictive control \cite{garcia1989model,hovland2008explicit} that require multiple evaluations of a model for multiple inputs. \textcolor{rev1}{Proper orthogonal decomposition (POD) \cite{holmes1998turbulence} and dynamic mode decomposition \cite{kutz2016dynamic} enabled ROMs are some of the most widely used methods among the fluid dynamics community. POD is often combined with Galerkin projection (GP) to model the dynamics of the ROM \cite{rempfer2000low,rowley2004model}. One of the limitations of intrusive approaches like GP is that it requires the complete knowledge about the equations governing the system's dynamics. However, for many physical systems, the exact governing equations are not available due to imperfect knowledge about the system. For example, in many geophysical flows we might not have accurate models for processes such as wind forcing, bottom friction, stratification, and other parameterization \cite{krasnopolsky2006complex}.}

Lately, equation-free or non-intrusive reduced order modeling (NIROM) has drawn great attention in the fluid mechanics' community due to its flexibility and efficiency for systems with incomplete dynamics and is undergoing rapid development with various algorithms emerging from different perspectives \cite{yu2019non}. \textcolor{rev1}{In most of the NIROMs, the main idea is to employ deep learning methods to construct nonlinear manifold \cite{gonzalez2018deep,lee2020model,maulik2021reduced} and to model the temporal dynamics of ROMs \cite{xu2020multi,rahman2019nonintrusive,wang2018model,guo2019data}.} Despite the success of NIROMs for many complex nonlinear problems, the naive deployment of NIROMs in multi-query applications is limited because they lack connection with the physics-based model and might suffer from poor performance in extrapolation regime \cite{lui_wolf_2019}. Therefore, the hybrid modeling approach that can overcome these drawbacks is warranted and to this end, we apply the PGML framework that exploits the Galerkin projection model for the known physics and long-short term memory (LSTM) neural network to model the unknown physics. \textcolor{rev1}{We remark here that the ideas from other approaches like  physics-reinforced neural network (PRNN) \cite{Chen2020PhysicsinformedML} and physics-embedded convolutional autoencoder (PhyCAE) \cite{mohan2020embedding} can be easily integrated within the proposed PGML framework.} 

In this study, we consider a dynamical system whose evolution is governed by partial differential equations such as the Navier-Stokes equations as follows 
\begin{equation} \label{eq:dyn_system0} 
    \dot{\boldsymbol{u}} = \boldsymbol{f}(\boldsymbol{u};\boldsymbol{x},t;\boldsymbol{\mu}), 
\end{equation}
where $\boldsymbol{u}$ is the prognostic variable that depends on a set of parameters, $\boldsymbol{f}$ is the nonlinear function that models the known processes and conservation laws, and $\boldsymbol{\mu} \in \mathbb{R}^{N_p}$ is the parameterization such as Reynolds or Rayleigh numbers. However, there could be unknown physics that might not be covered within the physics-based model $\boldsymbol{f}$. The augmented representation can then be written at an abstract level as
\begin{equation} \label{eq:dyn_system} 
    \dot{\boldsymbol{u}} = \boldsymbol{f}(\boldsymbol{u};\boldsymbol{x},t;\boldsymbol{\mu}) + \boldsymbol{\pi}(\boldsymbol{u};\boldsymbol{x},t;\boldsymbol{\mu}), 
\end{equation}
where $\boldsymbol{\pi}$ encompasses all the unknown processes and parameterizations. Here, we can define $\delta \boldsymbol{u}$ as a natural way to quantify the unknown physics $\boldsymbol{\pi}$ as follows
\begin{equation} \label{eq:norm}
\delta \boldsymbol{u} = \sqrt{(|| \boldsymbol{u}_1 || - || \boldsymbol{u}_2 ||)^2} / \sqrt{|| \boldsymbol{u}_1 ||^2 + || \boldsymbol{u}_2 ||^2},
\end{equation}
where $\boldsymbol{u}_1$ is the solution of Eq.~\ref{eq:dyn_system} without $\boldsymbol{\pi}$, $\boldsymbol{u}_2$ is the solution of Eq.~\ref{eq:dyn_system} with $\boldsymbol{\pi}$, and $|| \cdot ||$ is the $L_2$ norm. One can easily quantify the level of unknown processes using the definition given by Eq.~\ref{eq:norm}, i.e. $\delta\boldsymbol{u} \rightarrow 0$ when $\boldsymbol{\pi} \rightarrow 0$. On the other hand, a higher value of $\delta\boldsymbol{u}$, i.e. close to 1.0, means that our known physical model $\boldsymbol{f}$ is poor. 

\textcolor{rev1}{The chief motivation behind the PGML framework is to exploit domain knowledge to produce generalizable data-driven models.} We demonstrate the application of the PGML framework to fuse intrusive and non-intrusive parametric projection-based ROMs for incomplete dynamical systems that can be represented using Eq.~\ref{eq:dyn_system}.  Apart from improving the extrapolation performance of data-driven models, the PGML framework can also be considered as an enabler for multi-fidelity computing through a synergistic combination of low-cost/low-fidelity models with the high-fidelity data. The field of multi-fidelity computing is gaining attention in computational science and engineering, as it provides a way to utilize computationally cheap approximate models with high-fidelity observations \cite{peherstorfer2018survey,perdikaris2017nonlinear}. This is particularly relevant to fluid mechanics applications where there is a wealth of simplified models like potential flow theory, Blasius boundary layer model, and continuum Darcy/pipe flow model. One of the major challenges in multi-fidelity computing is determining the correlation between low and high fidelity models, and the neural networks employed within the PGML framework is capable of discovering this nonlinear relation. This first step in the assessment of the new PGML model shows that the proposed hybrid reduced order modeling framework could represent a viable tool for emerging digital twin applications \cite{rasheed2020digital}. 

\textcolor{rev1}{The paper is structured as follows. We introduce the parametric ROM framework in Section~\ref{sec:rom}. We then discuss the a specific formulation of the PGML framework in Section~\ref{sec:pgml} for combining the domain knowledge with data. The numerical results for the two-dimensional vorticity transport equation are detailed in Section~\ref{sec:results}. Finally, Section~\ref{sec:conclusion} will present conclusions and ideas for the future work.}

\section{Parametric ROM framework} \label{sec:rom}
The first step in parametric ROMs is the dimensionality reduction. We choose proper orthogonal decomposition (POD) to compute the linear orthogonal basis functions. \textcolor{rev1}{POD is one of the most popular technique for dimensionality reduction in the fluid mechanics community \cite{sirovich1987turbulence,berkooz1993proper}. POD provides the linear basis functions that minimizes the error between the true data and its projection in the $L_2$ sense compared to all linear basis functions of the same dimension.} Given the $N_s$ snapshots of the data for a variable of interest in the physical system, we can form the matrix $\boldsymbol{A}$ as follows
\begin{equation}
    \boldsymbol{A} = \bigg[ \boldsymbol{u}^{(1)},\boldsymbol{u}^{(2)}, \dots, \boldsymbol{u}^{(N_s)}\bigg] \in \mathbb{R}^{N \times N_s}
\end{equation}
where $\boldsymbol{u}^{(k)} \in \mathbb{R}^N$ corresponds to individual snapshot of the solution in time. Then, we perform the singular value decomposition (SVD) as follows
\begin{equation}
    \boldsymbol{A} = \boldsymbol{W}\boldsymbol{\Sigma}\boldsymbol{V}^T = \sum_{k=1}^{N_s}\sigma_k \boldsymbol{w}_k \boldsymbol{v}_k^T,
\end{equation}
where $\boldsymbol{W}\in \mathbb{R}^{N \times N_s}$ is a matrix with orthonormal columns which represent the left-singular vectors, $\boldsymbol{V} \in \mathbb{R}^{N_s \times N_s}$ is matrix with orthonormal columns representing the right-singular vectors, and $\boldsymbol{\Sigma} \in \mathbb{R}^{N_s \times N_s}$ is a matrix with positive diagonal entries, called singular values, arranged such that $\sigma_1 \ge \sigma_2 \ge \dots \ge \sigma_{N_s} \ge 0$. In dimensionality reduction, only first $N_r$ columns of $\boldsymbol{W}$ and $\boldsymbol{V}$ are retained along with the upper-left $N_r \times N_r$ sub-matrix of $\boldsymbol{\Sigma}$. \textcolor{rev1}{The singular values ${\sigma_i}$ give a measure of the quality of information that is retailed in $N_r$-order approximation of the matrix $\boldsymbol{A}$.} The $N_r$ columns of $\boldsymbol{W}$ are referred to as the POD modes or basis functions, denoted as $\boldsymbol{\Phi}_{\boldsymbol{\mu}} = \{ \phi_k\}_{k=1}^{N_r}$. \textcolor{rev1}{The amount of the energy retained by POD modes can be computed using the quantity called the relative information content (RIC) as follows
\begin{equation}\label{eq:ric}
    \text{RIC}(N_r) = \frac{\sum_{j=1}^{N_r} \sigma_j^2}{\sum_{j=1}^{N_s} \sigma_j^2}    
\end{equation}}

We utilize the Grassmann manifold interpolation \cite{amsallem2008interpolation,zimmermann2018geometric} to compute the basis functions for a new set of parameters from available POD basis function computed in the offline phase. \textcolor{rev1}{In the offline phase, we calculate a set of POD basis functions $\{\boldsymbol{\Phi}_i\}_{i=1}^{N_c}$ corresponding to a set of parameters $\{\boldsymbol{\mu}_i\}_{i=1}^{N_c}$. The Grassman manifold interpolation consists of choosing a reference point $S_{0}$ corresponding to the basis set $\mathbf{\Phi}_0$, and then mapping each point $S_i$ to a matrix $\mathbf{\Gamma}_i$ which represents a point on the tangent space at $S_{0}$ using logarithmic map $\text{Log}_{S_{0}}$ as follows
\begin{equation}
    (\boldsymbol{\Phi}_i-\boldsymbol{\Phi}_0 \boldsymbol{\Phi}_0^T \boldsymbol{\Phi}_i)(\boldsymbol{\Phi}_0^T\boldsymbol{\Phi}_i)^{-1} = \boldsymbol{W}_i\boldsymbol{\Sigma}_i \boldsymbol{V}_i^T,
\end{equation}
\begin{equation}
    \boldsymbol{\Gamma}_i =  \boldsymbol{W}_i \text{tan}^{-1}(\boldsymbol{\Sigma}_i) \boldsymbol{V}_i^T.
\end{equation}
The matrix $\boldsymbol{\Gamma}_t$ corresponding to target operating point $\boldsymbol{\mu}_t$ is 
computed by interpolating the corresponding entries of matrices $\{\mathbf{\Gamma}_i\}_{i=1}^{N_c}$. When $N_p=1$, typically Lagrange-type interpolation method is applied as follows
\begin{equation}
    \mathbf{\Gamma}_t = \sum_{i=1}^{N_c}\bigg( \prod_{\substack{j=1 \\ j\neq i}}^{N_c}\frac{\mu_t - \mu_j}{\mu_i - \mu_j}\bigg)\mathbf{\Gamma}_i.
\end{equation}
The POD basis functions $\boldsymbol{\Phi}_t$ corresponding to the test parameter $\boldsymbol{\mu}_t$ is computed using the exponential mapping as follows 
\begin{equation}
    \boldsymbol{\Gamma}_t = \boldsymbol{W}_t \boldsymbol{\Sigma}_t \boldsymbol{V}_{t}^{T},
\end{equation}
\begin{equation}
    \boldsymbol{\Phi}_t = [\boldsymbol{\Phi}_0 \boldsymbol{V}_t \text{cos}(\boldsymbol{\Sigma}_t) + \boldsymbol{W}_t \text{sin}(\boldsymbol{\Sigma}_t)]\boldsymbol{V}_{t}^{T},
\end{equation}
where the trigonometric operators apply only to diagonal elements.
}

Once the basis functions corresponding to the test operating point are computed, the reduced order representation of the high-dimensional state $\boldsymbol{u}(\boldsymbol{x},t;\boldsymbol{\mu})$ is
\begin{equation} \label{eq:reduced_state}
    \boldsymbol{u}(\boldsymbol{x},t;\boldsymbol{\mu}) \approx \boldsymbol{u}_r(\boldsymbol{x},t;\boldsymbol{\mu}) = \boldsymbol{\Phi}_{\boldsymbol{\mu}} \boldsymbol{a}(t;\boldsymbol{\mu}),
\end{equation}
where $\boldsymbol{a}(t;\boldsymbol{\mu}) \in \mathbb{R}^{N_r}$ is the reduced state on the basis functions and is also called as the modal coefficients and $\boldsymbol{\Phi}_{\boldsymbol{\mu}}$ is the POD basis functions for a set of parameters $\boldsymbol{\mu}$. The ROM is obtained by substituting the low-rank approximation given in Eq.~\ref{eq:reduced_state} into the full order model defined in Eq.~\ref{eq:dyn_system} and then taking the inner product with test basis functions to yield a system of $N_r$ ordinary differential equations (ODEs). For many fluid mechanics problems, the mechanistic description of some variables or processes is not available or insufficient for the desired task (i.e., the term $\boldsymbol{\pi}(\cdot)$ is unknown). Therefore, the physics-based intrusive approaches like Galerkin projection ROM (GROM) provide us
\begin{equation}
    \dot{\boldsymbol{a}} =  \boldsymbol{\Phi}_{\boldsymbol{\mu}}^T \boldsymbol{f}(\boldsymbol{\Phi}_{\boldsymbol{\mu}} \boldsymbol{a};\boldsymbol{x},t;\boldsymbol{\mu}).
\end{equation}
We note here that in GROM, the test basis functions are the same as the trial basis which allows us to make use of the orthonormality, i.e., $\boldsymbol{\Phi}_{\boldsymbol{\mu}}^T\boldsymbol{\Phi}_{\boldsymbol{\mu}} = \boldsymbol{I}_{N_r}$. 

\textcolor{rev1}{One of the limitation with intrusive ROMs like GROM is that the governing equations of the system have to be known exactly. If the governing equations are not known exactly, the effect of unknown dynamics is truncated in GROMs and accurate prediction is not achieved.} This issue is mitigated in NIROMs that exploit machine learning (ML) algorithms to learn the reduced order dynamics from the observed data \cite{yu2019non}. The training data for the ML algorithm can be generated by using the same data used for computing the POD basis functions as follows 
\begin{equation}
    \boldsymbol{\alpha}(t;\boldsymbol{\mu}) = \langle \boldsymbol{u}(\boldsymbol{x}, t; \boldsymbol{\mu}) , \boldsymbol{\Phi}_{\boldsymbol{\mu}} \rangle,
\end{equation}
where $\boldsymbol{\alpha}$ is the data projection coefficient, and the angle-parentheses refer to the Euclidean inner product defined as $\langle \boldsymbol{p} , \boldsymbol{q} \rangle = \boldsymbol{p}^T \boldsymbol{q}$. The neural network is one of the most successful algorithms for learning the dynamics on reduced order basis functions. However, as the complexity of the problems increases, the number of trainable parameters quickly explodes and neural networks suffer from high epistemic uncertainty associated with limited training data. We address this problem by hybridizing intrusive and non-intrusive approaches through our PGML framework. 
\vspace{-10pt}

\section{Physics-guided machine learning (PGML) framework} \label{sec:pgml}

\begin{figure*}
\centering
\mbox{\subfigure{\includegraphics[width=0.98\textwidth]{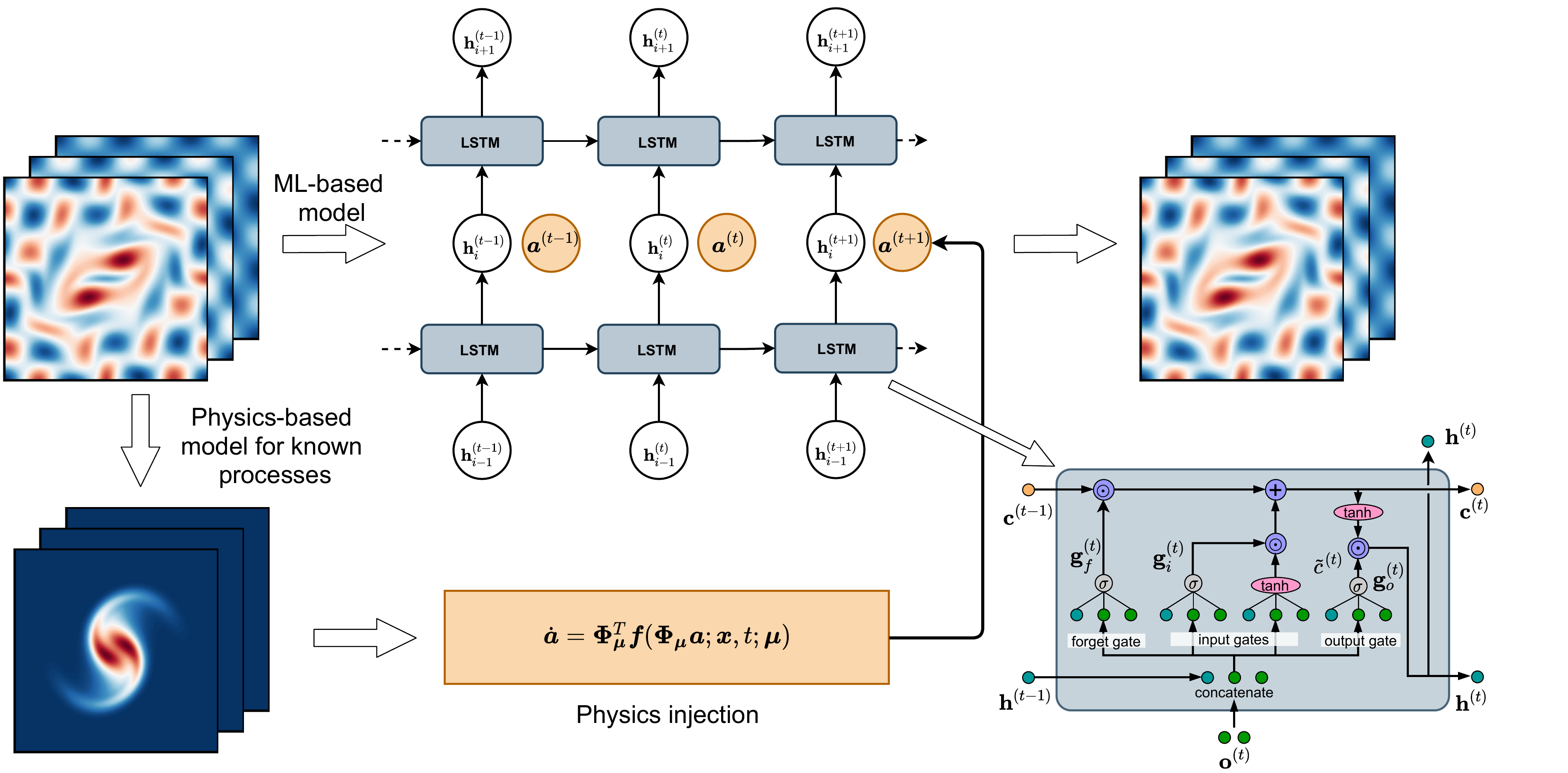}}}
\caption{Proposed physics-guided machine learning (PGML) framework for reduced order modeling, where the physics-based features are directly embedded into hidden layers of the neural network along with learned latent variables.} 
\label{fig:pgml}
\end{figure*}


In this Section, we detail different components of the proposed PGML framework for modeling the reduced order representation of the high-dimensional nonlinear dynamical system. \textcolor{rev1}{Different components of the PGML framework are illustrated in Figure~\ref{fig:pgml}.} First, we present the formulation of the LSTM neural network for time series modeling and then discuss its application within the PGML framework. In a typical ML algorithm for time series modeling, the model is trained on a time series of observable $\{\mathbf{o}^{(1)},\dots,\mathbf{o}^{(T)}\}$, where $\mathbf{o} \in \mathbb{R}^{d_o}$, sampled at a fixed time interval. The LSTM neural network is one of the most popular ML algorithms for time-series modeling due to its ability to model long-term temporal dependencies without suffering from the vanishing gradient problem of recurrent neural network \cite{hochreiter1997long}. \textcolor{rev1}{Recently, LSTM has also been very successful in modeling high-dimensional spatio-temporal chaotic systems and for ROMs \cite{pathak2018model, vlachas2018data, maulik2020non, vlachas2020backpropagation, hasegawa2020cnn}. The general functional form of the models used for time series forecasting can be written as
\begin{align}
    \mathbf{h}^{(t)} &= f_{h}^{h} (\mathbf{o}^{(t)}, \mathbf{h}^{(t-1)}), \\
    \tilde{\mathbf{o}}^{(t+1)} &= f_{h}^{o} (\mathbf{h}^{(t)}),
\end{align}
where $\mathbf{o}^{(t)}$ is the current observable and $\tilde{\mathbf{o}}^{(t+1)}$ is the forecast for the observable at the next time step $\mathbf{o}^{(t+1)}$ and $\mathbf{h}^{(t)} \in \mathbb{R}^{d_h}$ is the hidden state. Here. $f_h^h$ is the hidden-to-hidden mapping and $f_h^o$ is the hidden-to-output mapping. 
}

\textcolor{rev1}{
The LSTM mitigates the issue with vanishing (or exploding) gradient by employing the gating mechanism that allows information to be forgotten. The equations that implicitly define the hidden-to-hidden mapping from hidden state from the previous time step (i.e., $\mathbf{h}^{(t-1)}$) and input vector at the current time step (i.e., $\mathbf{o}^{(t)}$) to the forecast hidden state (i.e., $\mathbf{h}^{(t)}$) can be written as
\begin{align}
    \mathbf{g}^{(t)}_f &= \sigma_f(\mathbf{W}_f [\mathbf{h}^{(t-1)},\mathbf{o}^{(t)}] + \mathbf{b}_f), \\
    \mathbf{g}^{(t)}_i &= \sigma_i(\mathbf{W}_i [\mathbf{h}^{(t-1)},\mathbf{o}^{(t)}] + \mathbf{b}_i), \\
    \tilde{\mathbf{c}}^{(t)} &= \text{tanh}(\mathbf{W}_c [\mathbf{h}^{(t-1)},\mathbf{o}^{(t)}] + \mathbf{b}_c), \\
    \mathbf{c}^{(t)} &= \mathbf{g}^{(t)}_f \odot \mathbf{c}^{(t-1)} + \mathbf{g}^{(t)}_i \odot \tilde{\mathbf{c}}^{(t)}, \\
    \mathbf{g}^{(t)}_o &= \sigma_o(\mathbf{W}_o [\mathbf{h}^{(t-1)},\mathbf{o}^{(t)}] + \mathbf{b}_o), \\
    \mathbf{h}^{(t)} &= \mathbf{g}^{(t)}_o \odot \text{tanh} (\mathbf{c}^{(t)}),
\end{align}
where $\mathbf{g}^{(t)}_f,~\mathbf{g}^{(t)}_i,~\mathbf{g}^{(t)}_o \in \mathbb{R}^{d_h}$ are the forget gate, input gate, and output gate, respectively. The $\mathbf{o}^{(t)} \in \mathbb{R}^{d_o}$ is the input vector at time $t$, $\mathbf{h}^{(t)} \in \mathbb{R}^{d_h}$ is the hidden state, $\mathbf{c}^{(t)} \in \mathbb{R}^{d_h}$ is the cell state, $\mathbf{W}_f,~\mathbf{W}_i,~\mathbf{W}_c,~\mathbf{W}_o \in \mathbb{R}^{d_h \times (d_h + d_o)}$ are the weight matrices, and $\mathbf{b}_f,~\mathbf{b}_i,~\mathbf{b}_c,~\mathbf{b}_o \in \mathbb{R}^{d_h}$ are the bias vectors. The symbol $\odot$ denotes the element-wise multiplication, and $\sigma$ is the sigmoid activation function. The activation functions $\sigma_f, \sigma_i, \sigma_o$ are sigmoids. The hidden-to-output mapping is given by a fully connected layer with a linear activation function as follows 
\begin{equation}
    \tilde{\mathbf{o}}^{(t+1)} = \mathbf{W}_o \mathbf{h}^{t},
\end{equation}
where $\mathbf{W}_o \in \mathbb{R}^{d_o \times d_h}$.
}

When we apply the LSTM neural network for building NIROMs, the reduced order state of the system at a future time step, i.e., $\boldsymbol{\alpha}^{(t+1)}$, is learned as the function of a short history of $d$ past temporally consecutive reduced order states as follows 
\begin{equation}
    \boldsymbol{\alpha}^{(t+1)} = \mathcal{F}(\underbrace{\boldsymbol{z}^{(t)}, \boldsymbol{z}^{(t-1)}, \dots , \boldsymbol{z}^{(t-d+1)}}_{\boldsymbol{z}^{(t):(t-d+1)}}; \boldsymbol{\theta}),
\end{equation}
where $\mathcal{F}(\cdot ~ ; \boldsymbol{\theta})$ is the nonlinear function parameterized by a set of parameters $\boldsymbol{\theta}$, and $\boldsymbol{z}$ refers to input features consisting of the POD modal coefficients and a set of parameters governing the system, i.e., $\boldsymbol{z} \in \mathbb{R}^{N_r+N_p}$. The LSTM is trained using the backpropagation through time (BPTT)  algorithm and the parameters are optimized as
\begin{equation}
    \boldsymbol{\theta}^{*} = \argmin_{\theta} \frac{1}{N-d+1} \sum_{n=d}^N ||\mathcal{F}(\underbrace{\boldsymbol{z}^{(t):(t-d+1)}}_{\zeta}; \boldsymbol{\theta}) - \boldsymbol{\alpha}^{(t+1)}||_2^2.
\end{equation}
We employ $(N_l-1)$ LSTM layers and a single dense layer with a linear activation function as the last layer. Therefore, the ML model can be written as
\begin{equation}
    \mathcal{F}_{\text{ML}}(\boldsymbol{\zeta};\boldsymbol{\theta}) = \boldsymbol{h}_{N_l}(\cdot ; \boldsymbol{\Theta}_{N_l}) \circ \cdots \circ \boldsymbol{h}_{2}(\cdot ; \boldsymbol{\Theta}_{2}) \circ \boldsymbol{h}_{1}(\boldsymbol{\zeta} ; \boldsymbol{\Theta}_{1}),
\end{equation}
where the output of each LSTM layer ($i=1,\dots, N_l-1$) is $\boldsymbol{h}_{i}(\cdot ; \boldsymbol{\Theta}_{i}) \in \mathbb{R}^{d \times N_{h}}$ and the last dense layer maps the final hidden state to the output, i.e., $\boldsymbol{h}_{N_l}(\cdot ; \boldsymbol{\Theta}_{N_l}): \mathbb{R}^{N_h} \rightarrow \mathbb{R}^{N_r}$. Here, $N_h$ is the number of hidden units in the LSTM cell and we use the constant number of hidden units across all LSTM layers. 

In the PGML framework, the features extracted from the physics-based model are embedded into the $i$th intermediate hidden layer along with the latent variables as follows
\begin{equation}
    \mathcal{F}_{\text{PGML}}(\boldsymbol{\zeta};\boldsymbol{\theta}) = \boldsymbol{h}_{N_l}(\cdot ; \boldsymbol{\Theta}_{N_l}) \circ \cdots     \circ \underbrace{\mathcal{C}(\boldsymbol{h}_{i}(\cdot ; \boldsymbol{\Theta}_{i}),\boldsymbol{a}^{(t):(t-d+1)})}_{\text{Physics injection}} \circ 
    \cdots \circ \boldsymbol{h}_{1}(\boldsymbol{\zeta} ; \boldsymbol{\Theta}_{1}),
\end{equation}
where $\mathcal{C(\cdot,\cdot)}$ represents the concatenation operation and $\mathbf{a}$ are the physics-based features (i.e., the GROM modal coefficients). Therefore, the output of $i$th layer will be in $\mathbb{R}^{d \times (N_r+N_h)}$. \textcolor{rev1}{We highlight here that the physics-based features are embedded only at a single layer of the neural network. Therefore, at which layer shall the physics-based features be embedded becomes another hyperparameter of the PGML framework. While some guidelines or rule of thumb can be established based on the interplay between known and unknown physics, a thorough systematic investigation is needed. Methods like layer-wise relevance propagation can be adopted to bring interpretability to complex neural networks in the PGML framework and we consider it as part of our future work.} The physics-based features assist the LSTM network in constraining the output to a manifold of the physically realizable solution and leads to improved generalizability for the extrapolation regime. Even though we demonstrate the PGML framework for POD-ROM, we emphasize here that the PGML framework is highly modular and can also be applied to nonlinear, cluster-based, and network-based ROMs. \textcolor{rev1}{For example, we do not need to restrict to only linear basis construction methods like POD and the PGML framework can be easily extended for nonlinear manifold generation methods like convolutional autoencoders \cite{gonzalez2018deep,lee2020model,eivazi2020deep}.}

\vspace{-10pt}

\section{Vortex merging experiments} \label{sec:results}

\begin{figure}
\centering
\mbox{\subfigure{\includegraphics[width=0.45\textwidth]{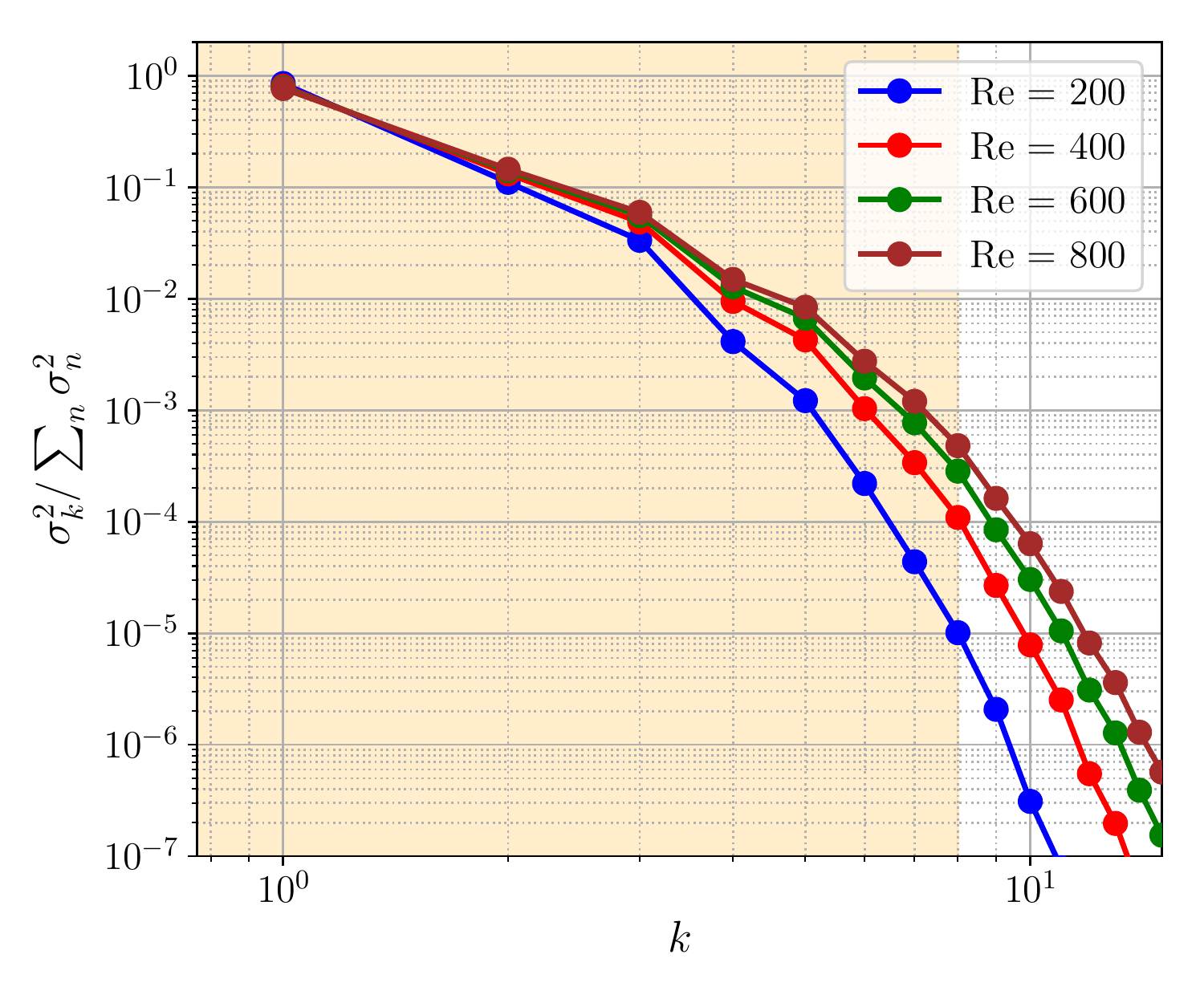}}}
\caption{Amount of the kinetic energy captured by each POD modes.} 
\label{fig:ric}
\end{figure}

\begin{figure*}
\centering
\mbox{\subfigure{\includegraphics[width=0.95\textwidth]{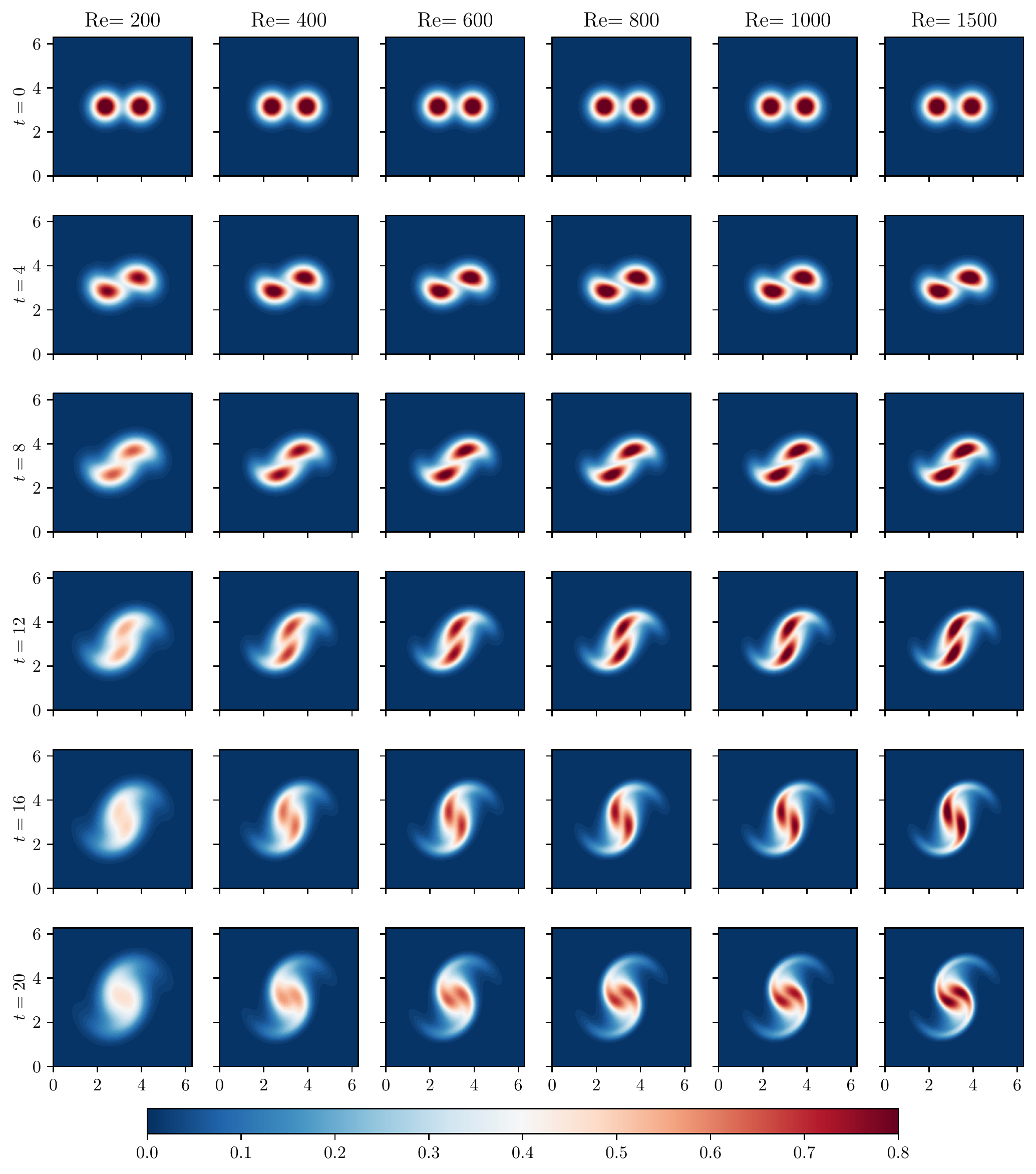}}}
\caption{Temporal evolution of the vorticity field at different time instances for several Reynolds number investigated in this study. We note here that the vorticity field is presented for $\gamma=0.0$.} \vspace{-10pt}
\label{fig:vort_field}
\end{figure*}

We apply the PGML framework to the two-dimensional vorticity transport equation for the vortex merger problem as a prototypical test case. This problem involves the study of the merging of two co-rotating vortices. The two-dimensional vorticity transport equation can be written as 



\begin{equation} \label{eq:ns2d}
\dfrac{\partial \omega}{\partial t} + J(\omega, \psi)=   \dfrac{1}{\text{Re}} \nabla^2 \omega + \boldsymbol{\pi}(\boldsymbol{x},t;\boldsymbol{\mu}),
\end{equation}

\begin{align} \label{eq:ns2d-poisson}
\nabla^2 \psi = \omega, 
\end{align}
where $\omega$ is the vorticity defined as $\omega = \nabla \times \mathbf{u}$, $\mathbf{u} = [u,v]^T$ is the velocity vector, $\psi$ is the streamfunction, $\text{Re}$ is the Reynolds number parameterizing the system and $\boldsymbol{\pi}(\cdot)$ represent the unknown physics. \textcolor{rev1}{We can rewrite Eq.~\ref{eq:ns2d} as
\begin{equation} \label{eq:ns2d1}
\dfrac{\partial \omega}{\partial t} = \dfrac{1}{\text{Re}} \nabla^2 \omega - J(\omega, \psi) + \boldsymbol{\pi}(\boldsymbol{x},t;\boldsymbol{\mu}).
\end{equation}}
\textcolor{rev1}{The first two terms on the right hand side of Eq.~\ref{eq:ns2d1}, i.e., the linear Laplacian term and the nonlinear Jacobian term represent the known physics of the system and are defined as follows
\begin{align}
    \nabla^2 \omega &= \dfrac{\partial^2 \omega}{\partial x^2} + \dfrac{\partial^2 \omega}{\partial y^2}, \\
    J(\omega, \psi) &= \dfrac{\partial \psi}{\partial y}\dfrac{\partial \omega}{\partial x} - \dfrac{\partial \psi}{\partial x} \dfrac{\partial \omega}{\partial y}.
\end{align}}
\textcolor{rev1}{As discussed in Section~\ref{sec:rom}, the reduced order approximation of the vorticity and streamfunction field can be written as
\begin{align}
    \omega(x,y,t) \approx \sum_{k=1}^{N_r}a_k(t)\phi_k^{\omega}(x,y), \label{eq:omega_rom}\\
    \psi(x,y,t) \approx \sum_{k=1}^{N_r}a_k(t)\phi_k^{\psi}(x,y), \label{eq:psi_rom}
\end{align}}
\textcolor{rev1}{We have omitted the parameter $\boldsymbol{\mu}$ in Eq.~\ref{eq:omega_rom}-\ref{eq:psi_rom} for simplified notation. The vorticity and streamfunction share the same time dependent modal coefficients as they are both related by the kinematic relationship given in Eq.~\ref{eq:ns2d-poisson}. Furthermore, the POD basis functions of the streamfunction can be obtained by solving the below Poisson equation
\begin{equation}
    \nabla^2 \phi_k^{\psi} = \phi_k^{\omega}.
\end{equation}}
\textcolor{rev1}{The GROM for the vorticity transport equations considering only the known physics can be written as
\begin{equation} \label{eq:gp0}
     \boldsymbol{\dot{a}} =  {\mathfrak{L}} \boldsymbol{a} + \boldsymbol{a}^T {\mathfrak{N}} \boldsymbol{a},
\end{equation}
or more explicitly,
\begin{equation} \label{eq:gp}
    \dfrac{\text{d}a_k}{\text{d}t} =  {\sum_{i=1}^{N_r} \mathfrak{L}^{i}_{k} a_i + \sum_{i=1}^{N_r} \sum_{j=1}^{N_r} \mathfrak{N}^{ij}_{k} a_i a_j} ,
\end{equation}
where $\mathfrak{L}$ and $\mathfrak{N}$ are the linear and nonlinear operators. The linear and nonlinear operators for the vorticity transport equation are as follows
\begin{align}
    \mathfrak{L}_{ik} &= \bigg \langle \dfrac{1}{\text{Re}} \nabla^2 \phi_i^{\omega},\phi_k^{\omega} \bigg \rangle, \\
    \mathfrak{N}_{ijk} &= \bigg \langle -J(\phi_i^{\omega}, \phi_j^{\psi}),\phi_k^{\omega} \bigg \rangle,
\end{align}
where the angle-parentheses refer to the Euclidean inner product defined as $\langle \mathbf{x} , \mathbf{y} \rangle = \mathbf{x}^T \mathbf{y} =  \sum_{i=1}^{N} x_i y_i$. We apply the third-order Adams-Bashforth (AB3) numerical scheme to solve the system of ODEs given in Eq.~\ref{eq:gp}. In discrete sense, the update formula can be written as 
\begin{equation}
    {a}_k^{(n+1)} = {a}_k^{(n)} + {\Delta t} \sum_{q=0}^{s} \beta_q G({a}_k^{(n-q)}),   
\end{equation}
where $s$ and $\beta_q$ are the constants corresponding to AB3 scheme, which are $s=2,~ \beta_0=23/12,~ \beta_1=-16/12,$ and $\beta_2=5/12$. Here, the operator $G({a}_k)$ is the right hand side of Eq.~\ref{eq:gp} as follows
\begin{equation}
    G(a_k) = {\sum_{i=1}^{N_r} \mathfrak{L}^{i}_{k} a_i + \sum_{i=1}^{N_r} \sum_{j=1}^{N_r} \mathfrak{N}^{ij}_{k} a_i a_j}.
\end{equation}
}
\textcolor{rev1}{We start with an initial vorticity field of two Gaussian-distributed vortices,
\begin{align}
    \omega(x, y, 0) &= \exp\left( -\pi \left[ (x-x_1)^2  + (y-y_1)^2 \right] \right) \nonumber \\ &+ \exp{\left( -\pi \left[ (x-x_2)^2 + (y-y_2)^2 \right] \right)},
\end{align}
where their centers are initially located at $(x_1,y_1) = (3\pi/4,\pi)$ and $(x_2,y_2) = (5\pi/4,\pi)$.} We utilize an arbitrary array of decaying Taylor-Green vortices as the unknown processes parameterized by ${\mu} = \{\text{Re}\}$ i.e., we have 
\begin{equation}\label{eq:source_term}
    \boldsymbol{\pi}(x,y,t;\text{Re}) = -\gamma~ e^{-t/\text{Re}} \text{cos}(3 x) \text{cos}(3 y).
\end{equation}
The parameter $\gamma$ controls the strength of unknown physics compared to the known physics. The synthetic data for building the ROM is generated by solving Eq.~\ref{eq:ns2d} on the computational domain $(x,y) \in [0,2\pi]$ with periodic boundary conditions and $256^2$ spatial discretization. The system is evolved for 20 time units with a time-step size of $\Delta t = 1 \times 10^{-3}$. The data snapshots for the training are obtained for $\text{Re}=\{ 200,400,600,800 \}$ and the trained models are evaluated for $\text{Re}=\{ 1000,1500 \}$. \textcolor{rev1}{Specifically, the physical variable of interest is the vorticity field, i.e., $\boldsymbol{u}=\{\omega\}$ and the unknown source term $\boldsymbol{\pi}(\cdot)$ is given in Eq.~\ref{eq:source_term}.} We retain $N_r=8$ POD modes for all Reynolds numbers and these POD modes captures more than 99\% of the energy of the system \textcolor{rev1}{as illustrated in Fig.~\ref{fig:ric}. Instantaneous snapshots of the vorticity field for different Reynolds numbers at different time instances are displayed in Fig.~\ref{fig:vort_field}. We note here that the vorticity field shown in Fig.~\ref{fig:vort_field} is obtained by solving the vorticity transport equation assuming that there is no unknown physics, i.e., setting $\gamma=0.0$.} 

The LSTM network in the ML model consists of $N_l=5$ layers with the first four layers as the LSTM layers and the dense layer with a linear activation function as the output layer. Each LSTM layer has $N_h=80$ hidden units and tanh activation function is applied. \textcolor{rev1}{From our experimentation with the hyperparameters, we found that the results were not improving by employing deeper network architecture and in some cases, deeper networks led to poor performance due to overfitting.} We utilize $d=3$, i.e., the state of the system for three previous time steps are used to predict the future state. \textcolor{rev1}{Since we are using the information of only $d$ past temporally consecutive states as the input, the LSTM can capture dependencies up to $d$ previous time steps. During the online deployment, the initial condition for the first $d$ time steps is provided. This information is used to predict the forecast state at $(d+1)$th time step. Then the state of the system from $2-(d+1)$ is used to predict the forecast state at $(d+2)$th time step. This procedure is continued until the final time step.} The PGML framework also uses the same LSTM network architecture and the features from the Galerkin projection model (i.e., GROM modal coefficients) are embedded into the third LSTM layer. \textcolor{rev1}{We reiterate here that the layer at which the physics based features should be embedded is another hyperparameter of PGML framework and methods like neural architecture search can be used for complex problems \cite{maulik2020recurrent}. The online deployment of the PGML framework is similar to the ML framework.} We characterize the generalizability of the ML and PGML model through uncertainty estimate computed using the deep ensembles method where ensemble of neural networks are trained with different initialization of parameters \cite{tibshirani1996comparison,heskes1997practical,lakshminarayanan2016simple}. \textcolor{rev1}{In deep ensembles method, the prediction from the ensemble of neural networks is interpreted as the epistemic uncertainty. Despite the simplicity of this method, it is appealing due to its scalability and strong empirical results showing that the uncertainty estimate is as good as the Bayesian neural networks \cite{lakshminarayanan2016simple}.} The quantitative performance for ML and PGML framework is measured through the statistics of the weighted root mean squared error (RMSE) defined as 

\begin{equation} \label{eq:wrmse}
    \epsilon(i) = \sum_{k=1}^{N_r} \lambda_k \bigg({\frac{1}{N_t}\sum_{n=1}^{N_t} (\alpha_k^{(n)} - \tilde{\alpha}_k^{(n)}(i))^2}\bigg)^{1/2},
\end{equation}
where $\lambda_k = {\sigma_k^2}/({\sum_{j=1}^{N_r} \sigma_j^2})$ represents the energy in each POD mode and $\tilde{\alpha}_k(i)$ is the prediction of the modal coefficients for $i$th ensemble. Each neural network in the ensemble set is initiated by using a different initial seed. The statistics of the weighted RMSE, i.e., the mean and the standard deviation are computed as follows
\begin{equation} \label{eq:rmse_stats}
    \bar{\epsilon} = \frac{1}{N_e}\sum_{i=1}^{N_e} \epsilon(i), \quad s_\epsilon = \bigg({\frac{\sum_{i=1}^{N_e} (\epsilon(i) - \bar{\epsilon})^2}{N_e-1}}\bigg)^{1/2},
\end{equation}
where $N_e$ is the size of ensemble of the neural network. We set $N_e=30$ for both ML and PGML models in our study. 

\begin{figure*}
\centering
\mbox{\subfigure{\includegraphics[width=0.95\textwidth]{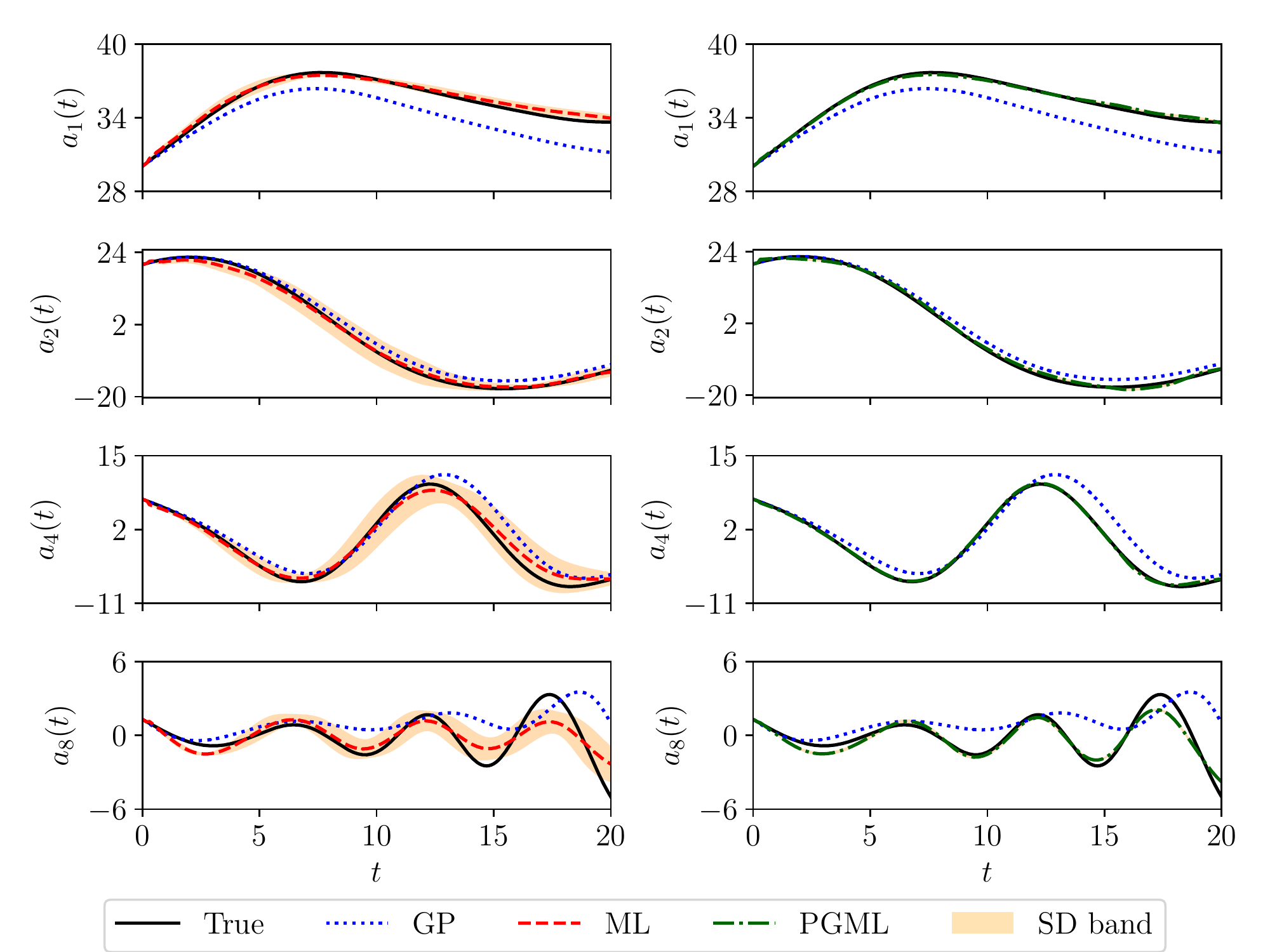}}}
\caption{Temporal evolution of selected modal coefficients for the vortex-merger test case at Re = 1500 and $\gamma = 0.01$. The ML (left) and PGML (right) represent the average of the modal coefficients predicted by all neural networks trained using MSE loss function with different seeds for initialization of the parameters. Although the ensemble-averaged ML model provides more accurate predictions than the GP model, there is a large standard deviation (SD) band over all ML model predictions. } \vspace{-10pt}
\label{fig:ml_pgml_modes}
\end{figure*}

\begin{figure*}
\centering
\mbox{\subfigure{\includegraphics[width=0.92\textwidth]{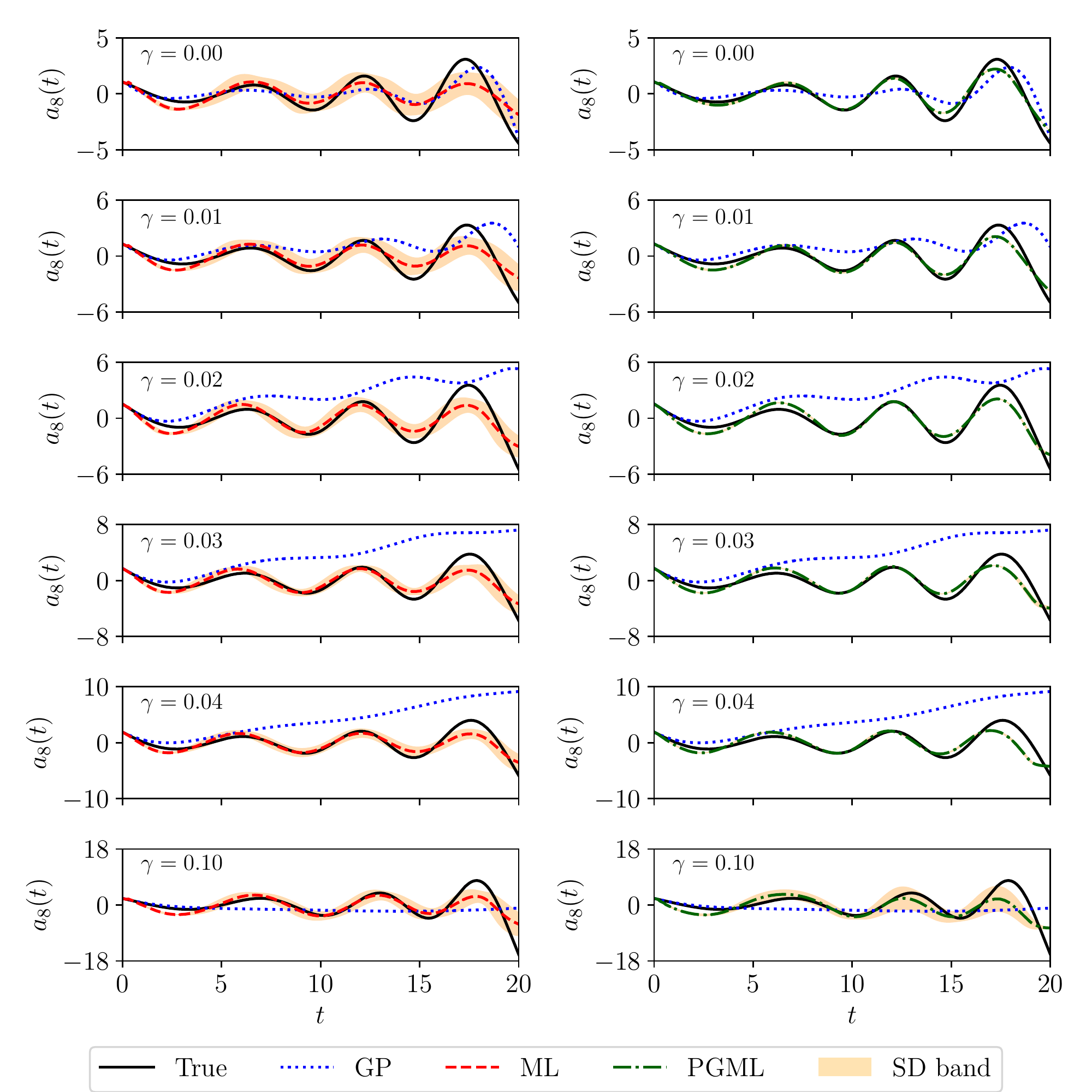}}}
\caption{Temporal evolution of the last modal coefficient for the vortex-merger test case at Re = 1500 and for different magnitudes of the source term. The dashed red curve represent the average of the modal coefficients predicted by ML model (left), the PGML model (right).}  \vspace{-5pt}
\label{fig:a8_modes}
\end{figure*}

Fig.~\ref{fig:ml_pgml_modes} shows the temporal evolution of selected modal coefficients predicted by true projection, GP, ML, and PGML models at $\text{Re}=1500$ for $\gamma = 0.01$. The ensemble-averaged modal coefficients predicted by the PGML model matches very accurately with the true projection compared to the GP and ML model. We also observe that the ensemble-average prediction by the ML model is also improved compared to the GP model. However, the prediction of the ML model is marred by high uncertainty compared to the PGML model prediction. 
\textcolor{rev1}{From Fig.~\ref{fig:vort_field}, we can notice that the flow is topologically very similar at different Reynolds numbers and the main difference is in the amplitude of the vorticity field. Therefore, the task of extrapolation for this problem is not very challenging for neural networks. However, in this study, we demonstrate the proof of concept to show the benefit of PGML over a purely data-driven ML model, and evaluating the PGML framework for topologically different data sets is a part of our future work.}
In Fig.~\ref{fig:a8_modes}, the temporal evolution of the last mode (i.e., $a_8$) predicted by true projection, GP, ML, and PGML model is displayed for different magnitudes of the unknown source term. Overall, we can observe that the PGML framework can predict the modal coefficients with high confidence up to $\gamma = 0.04$ and the prediction for $\gamma=0.1$ is less certain. \textcolor{rev1}{One of the reasons behind inaccurate prediction at $\gamma=0.1$ can be due to a very poor correlation between the known physics in the GP model and the actual physics of the system.} 

\begin{figure*}
\centering
\mbox{\subfigure{\includegraphics[width=0.92\textwidth]{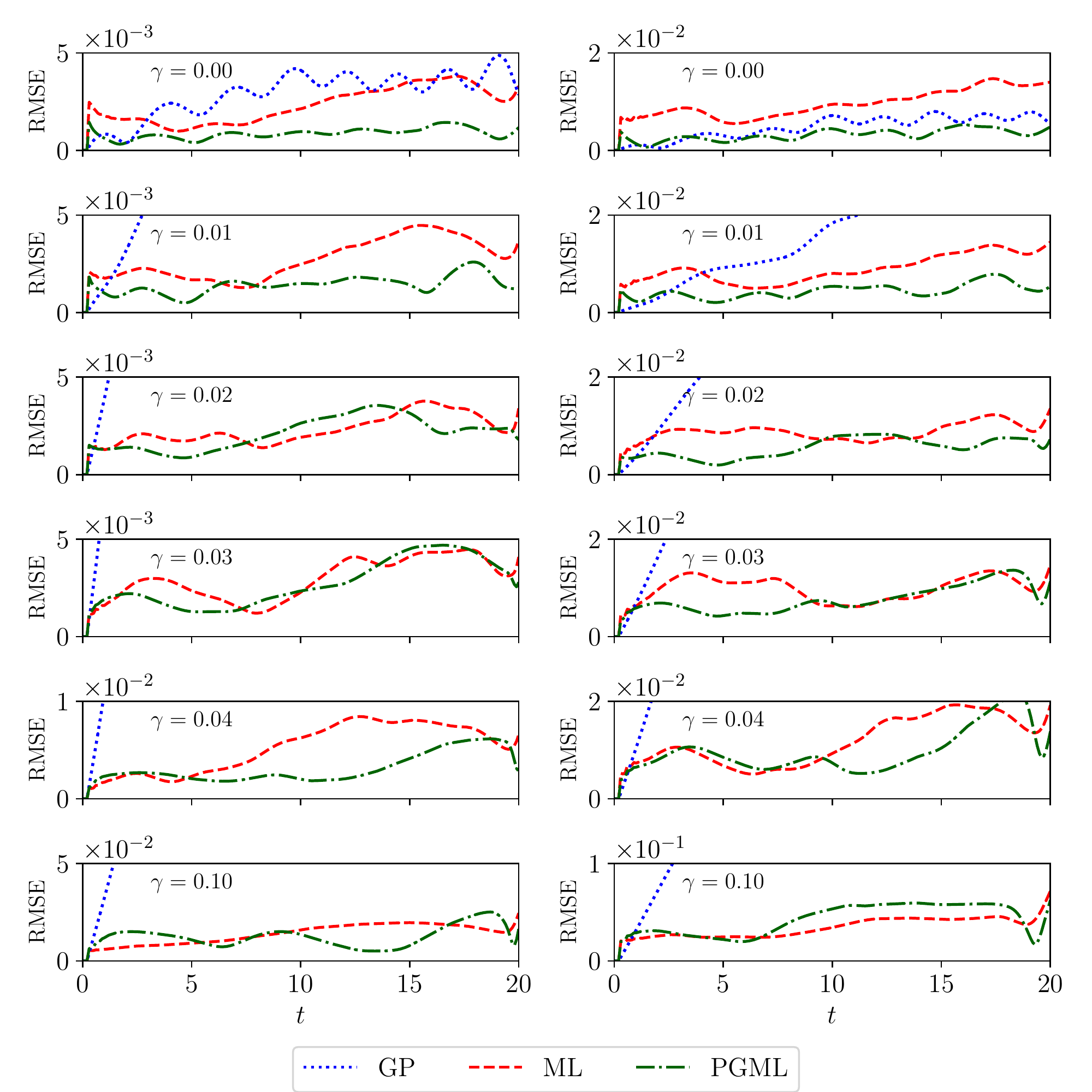}}}
\caption{Temporal evolution of the RMSE (defined in Eq.~\ref{eq:rmse}) for the vortex-merger test case at Re = 1000 (left) and Re = 1500 (right) for different magnitudes of the source term. The dashed red curve represent the average of the modal coefficients predicted by ML model (left), the PGML model (right). .}  \vspace{-5pt}
\label{fig:error_modes}
\end{figure*}

\begin{figure*}
\centering
\mbox{\subfigure{\includegraphics[width=0.95\textwidth]{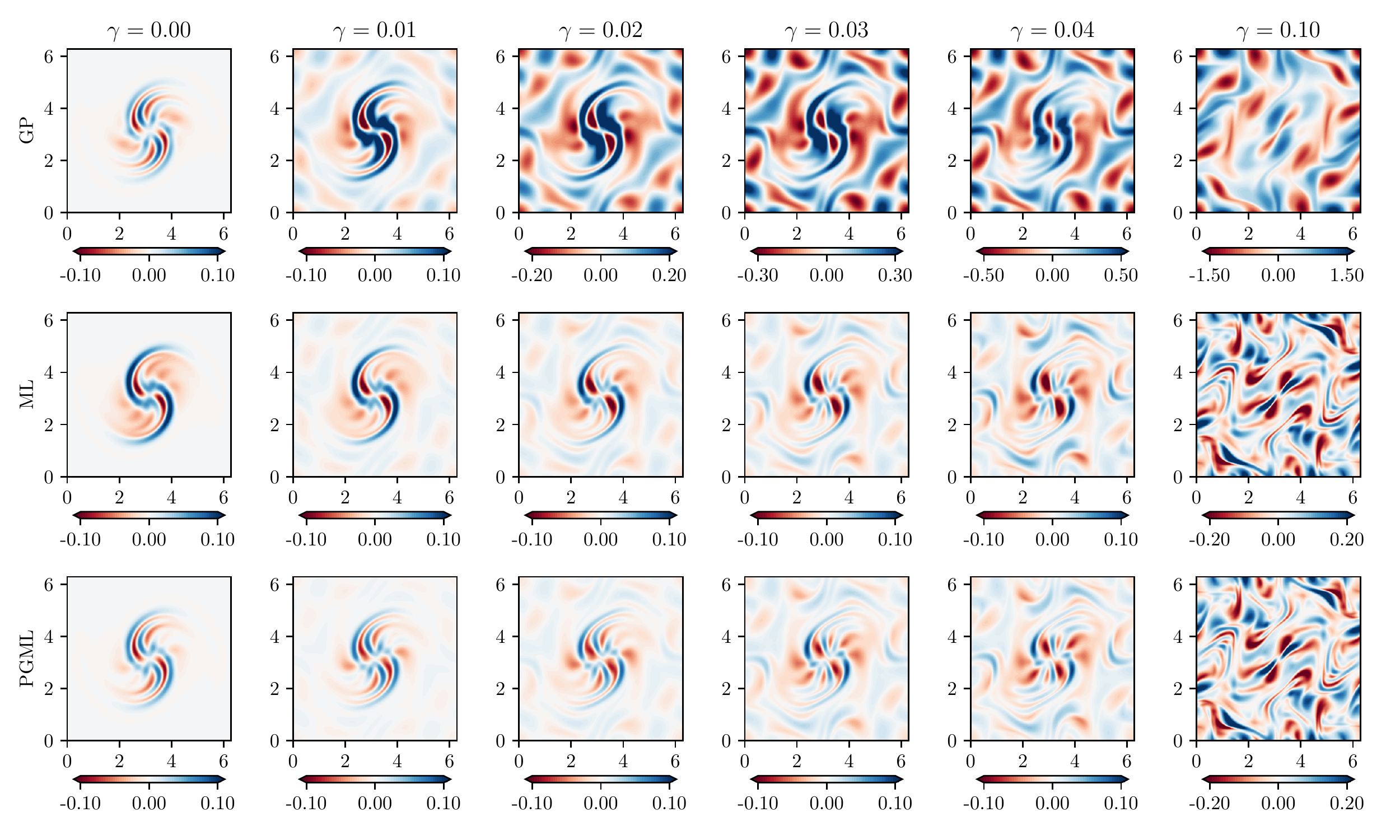}}}
\caption{The error in the prediction of the vorticity field at the final time $t=20$ for the GP model (top), ML model (middle), and PGML model (bottom).} \vspace{-5pt} 
\label{fig:error}
\end{figure*}

\textcolor{rev1}{We also evaluate the capability of GP, ML, and PGML framework in reconstructing the vorticity field using the root mean squared error (RMSE) between the true vorticity field and the predicted vorticity field. The RMSE is defined as
\begin{equation}\label{eq:rmse}
    \text{RMSE}(t) = \bigg(\frac{1}{N}\sum_{i=1}^{N}\bigg(\omega_{T}(\boldsymbol{x}_i,t) -  \omega_{R}(\boldsymbol{x}_i,t)\bigg)^2\bigg)^{1/2},
\end{equation}
where $N$ is the spatial resolution (i.e., $N=N_x \times N_y$), $\omega_T$ is the true projection of the vorticity field, and $\omega_R$ is the vorticity field predicted by ROM (computed using Eq.~\ref{eq:omega_rom}). The $\omega_R$ is computed using the ensemble-averaged modal coefficients predicted by ML and PGML model. Fig.~\ref{fig:error_modes} shows the time evolution of RMSE for different magnitudes of the source term at Re = 1000 and Re = 1500. We can observe that when the complete physics of the vortex-merging process is known (i.e., $\gamma=0.0$), the RMSE for both ML and PGML framework is less than the GP model at Re = 1000. However, at Re = 1500, the RMSE for the ML model is higher than the GP model. This is due to the poor extrapolation capability of a purely data-driven ML model. On the other hand, the PGML model can successfully use the physics-based information from the GP model leading to less RMSE compared to both GP and ML models. The vorticity field predicted by the GP model quickly diverges from the true vorticity field in the presence of unknown physics and overall the PGML prediction is better than the ML model as illustrated in Fig.~\ref{fig:error_modes}.}
The difference between the vorticity field at the final time $t=20$ predicted by true projection modal coefficients and GP, ML, and PGML models is depicted in Fig.~\ref{fig:error}.

Table~\ref{tab:rmse_weighted} reports the quantitative analysis of GP, ML, and PGML models at $\text{Re}=\{ 1000,1500 \}$ for different magnitudes of the unknown source term. The weighted RMSE mentioned in Table~\ref{tab:rmse_weighted} are calculated using Eq.~\ref{eq:wrmse} and its statistics is computed with Eq.~\ref{eq:rmse_stats}. Here, $\gamma = 0.0$ corresponds to the special case where the physics is completely available to us. In that case, there is no need to make a stochastic learning machine, so a physics-based model (e.g., GP) will provide an accurate prediction. \textcolor{rev1}{From Table~\ref{tab:rmse_weighted} we see that the weighted RMSE for the ML model is considerably higher for Re = 1500 at $\gamma=0.0$ due to the poor extrapolation ability of pure data-driven models. The weighted RMSE for the PGML model is slightly higher than the GP model even though the reconstructed vorticity field is more accurate as illustrated in Fig.~\ref{fig:error_modes}. The weighted RMSE assigns more weightage to the first few modes and therefore the inaccurate prediction for dominant modes leads to a higher value of weighted RMSE for the PGML model.} For higher values of $\gamma$, the dynamics of the system is not known exactly and it gives us an opportunity to introduce a stochastic neural network model. As demonstrated in Table~\ref{tab:rmse_weighted}, the trustworthiness of the pure data-driven ML model is less due to the high value of the uncertainty in the prediction (i.e., high standard deviation as shown in parenthesis). However, once we incorporate physics-based features from the GP model in the PGML framework, the uncertainty in the prediction is reduced by almost one order magnitude. We also consider the hypothetical situation where $\delta\boldsymbol{u}=0.6$ (so the physics is almost completely unknown and as shown in Figure~\ref{fig:error}, unknown physics severely dominates the vortex merging processes), then it is not surprising that the PGML framework would not be in favor of the standard ML because of the injection of the unrepresentative physics. Considering a false-negative analogy, this further verifies the success of the PGML methodology and will automatically indicate that the injected physics is not representative of the underlying processes. This feature of the PGML framework could act as a diagnostic approach in many digital twin applications to physical systems. 
There are numerous simulation packages specifically designed for modeling and computing $\boldsymbol{f}$ in these systems. On the other hand, exhaustive data sets become more and more available to generate versatile machine learning based predictive models. To this end, a confidence score $\delta\boldsymbol{u}$ defined in Eq.~\ref{eq:norm} can be utilized to assess the relative modeling strength of principled and data-driven models, where lower $\delta\boldsymbol{u}$ might indicate that the principled model is a reliable tool and the physics-based features from the principled model should be incorporated into data-driven models.


\begin{table*}
    \centering
    \begin{tabular}{p{0.05\textwidth}p{0.05\textwidth}p{0.06\textwidth}p{0.15\textwidth}p{0.15\textwidth} c p{0.07\textwidth}p{0.15\textwidth}p{0.15\textwidth}}
    \hline
    \multirow{2}{*}{$\gamma$} & \multirow{2}{*}{$\delta\boldsymbol{u}$} & \multicolumn{3}{c}{$\text{Re}=1000$} & & \multicolumn{3}{c}{$\text{Re}=1500$} \\ 
    \cline{3-5}  \cline{7-9} \\
     & & GP & ML & PGML & & GP & ML  & PGML \\
     \hline
    $ 0.00$ & 0.00 & 0.075	& 0.262 (0.190)	& 0.064 (0.012) & & 0.138 & 0.959 (0.632) & 0.192 (0.024) \\
    $ 0.01$ & 0.03 & 1.564	& 0.258 (0.194)	& 0.110 (0.012) & & 1.563 & 0.852 (0.521) & 0.300 (0.029) \\
    $ 0.02$ & 0.09 & 6.033	& 0.316	(0.113) & 0.181 (0.026) & & 6.061 & 0.998 (0.458) & 0.498 (0.040) \\
    $ 0.03$ & 0.17 & 12.809 & 0.465 (0.387) & 0.259 (0.040) & & 12.905 & 1.352 (0.582) & 0.784 (0.092) \\ 
    $ 0.04$ & 0.26 & 20.843 & 0.726 (0.591) & 0.312 (0.047) & & 21.082 & 1.828 (0.992) & 1.116 (0.149) \\
    $ 0.10$ & 0.60 & 70.935 & 1.910 (0.804) & 1.424 (0.356) & & 71.923 & 5.252 (2.387) & 5.842 (6.227) \\
    \hline
    \end{tabular}
    \caption{The mean of weighted root mean square error (and its standard deviation) in predicting the modal coefficients for higher Reynolds number flows. The training has been performed using data for $\text{Re} = \{200, 400, 600, 800\}$. Note that the control parameter $\gamma$ adjusts the relative strength between known-physics and unknown-physics (i.e., $\gamma=0$ refers to the special case when physics is fully known), and $\delta\boldsymbol{u}$ quantifies the level of unknown processes.} \vspace{-5pt}
    \label{tab:rmse_weighted}
\end{table*}

\vspace{-15pt}

\section{Discussion and conclusion} \label{sec:conclusion}

Although advanced machine learning (ML) models like deep learning networks are powerful tools for finding patterns in complicated datasets, the number of trainable parameters (weights) quickly explodes as these neural network models become complex, adversely affecting their interpretability and hence their trustworthiness. Our chief aim in this study is to illustrate how physics can be injected into these neural networks to improve the trustworthiness of the overall system. With this in mind, we introduce a physics-guided ML (PGML) framework that fuses first-principles with neural-network models, and explore complementary physics versus statistics issues that arise in the development of the PGML framework. \textcolor{rev1}{The PGML framework puts particular emphasis on the physics-based features and embeds them at an intermediate layer of the neural network.} The robustness of the PGML framework is successfully demonstrated for the vortex-merging process in the presence of an array of Taylor-Green vortices as the unknown source term. \textcolor{rev1}{Our results indicate that the PGML framework can give considerably accurate prediction in the extrapolation regime compared to its counterpart, i.e., pure data-driven ML model. The physics-based features from the GP model also ensures that the model uncertainty in the neural network prediction is reduced.} Finally, we also emphasize that the PGML framework is highly modular, and it naturally allows for the multi-fidelity model fusion in different branches of science and engineering.   

\textcolor{rev1}{The natural extension of the present study is to assess the PGML framework for much more complex flows, where there is a significant variation between the train and test data. Other unanswered questions are understanding how neural network assigns importance to physics-based features, at which layer shall the physics-based features be embedded, and methods from machine learning community can be utilized for this \cite{robinson2019dissecting,montavon2019layer}. Another direction for future work will be to use other uncertainty quantification methods like Monte Carlo dropouts, Bayesian neural networks to quantify model uncertainty.}

\section*{Acknowledgements}
This material is based upon work supported by the U.S. Department of Energy, Office of Science, Office of Advanced Scientific Computing Research under Award Number DE-SC0019290. O.S. gratefully acknowledges their support. 
Disclaimer: This report was prepared as an account of work sponsored by an agency of the United States Government. Neither the United States Government nor any agency thereof, nor any of their employees, makes any warranty, express or implied, or assumes any legal liability or responsibility for the accuracy, completeness, or usefulness of any information, apparatus, product, or process disclosed, or represents that its use would not infringe privately owned rights. Reference herein to any specific commercial product, process, or service by trade name, trademark, manufacturer, or otherwise does not necessarily constitute or imply its endorsement, recommendation, or favoring by the United States Government or any agency thereof. The views and opinions of authors expressed herein do not necessarily state or reflect those of the United States Government or any agency thereof.

\appendix


\bibliographystyle{unsrt} 
\bibliography{ref}

\end{document}